\def\np#1#2#3{{\it Nucl. Phys.\/} {\bf #1} (#2)#3}
\def\pl#1#2#3{{\it Phys. Lett.\/} {\bf #1} (#2)#3}
\def\prl#1#2#3{{\it Phys. Rev. Lett.\/} {\bf #1} (#2)#3}
\def\pr#1#2#3{{\it Phys. Rev.\/} {\bf #1} (#2)#3}

\def\zphys#1#2#3{{\it Z. Phys.\/} {\bf #1} (#2)#3}

\documentstyle[b_phys,12pt,epsf]{article}

\newcommand\gs{\stackrel{{\textstyle >}}{\sim}}
\newcommand\Dc{\Delta_D}
\newcommand\larr[1]{\raise1.5ex\hbox{$\leftarrow$}\mkern-16.5mu #1}
\newcommand\vslash{v\hskip-0.5em /}
\newcommand\overslash[1]{#1\hskip-0.6em /}
\newcommand\ie{{\it i.e.,\/}\ }
\newcommand\bas{{\alpha}_s}

\newcommand\eqn[1]{\begin{equation}#1\end{equation}}
\newcommand\eqna[1]{\begin{eqnarray}#1\end{eqnarray}}
\newcommand\vev[1]{\langle #1 \rangle}
\newcommand\Lambdab{\Lambda_b(v,s)}
\newcommand\Lambdac{\Lambda_c(v',s')}
\newcommand\ubar{\, {\overline u}(v',s')}
\newcommand\uvs{u(v,s)}
\newcommand\ccdot{\hbox{\kern-.1em$\cdot$\kern-.1em}}

\newcommand\vv{v \ccdot v'}
\newcommand\coupling{ {\bas(m_c) \over \pi}}
\newcommand\cLambdabar{{{\overline \Lambda} \over 2m_c}}
\newcommand\bLambdabar{{{\overline \Lambda} \over 2 m_b}}
\newcommand\fourthirds{{4 \over 3}}
\newcommand\g{\gamma}
\newcommand\CO{{\cal O}}
\newcommand\bqvqv[1]{{\overline Q'_{v'}#1 Q_v}}
\newcommand\opv{{\left({1+\vslash\over2}\right)}}
\newcommand\opvp{{\left({1+\vslash'\over2}\right)}}

\newcommand\ol{\overline}
\newcommand\Aslash{A\hskip-0.5em /}
\newcommand\im{{\rm i}}
\newcommand\Tr{{\rm Tr}}
\newcommand\del{\partial}
\newcommand\half{{\textstyle{1\over2}}} 
\newcommand\emnlk{\epsilon_{\mu\nu\lambda\kappa}}

\catcode`@=11
\let\@@@cite=\cite
\def\cite#1{{\egtrm\@@@cite{#1}}}

\def\@cite#1#2{%
   \let\@sf\empty
   \ifhmode\edef\@sf{
      \spacefactor\the\spacefactor}\/\fi        %
   $\relax{}^{#1\if@tempswa,#2\fi}$
   \@sf}

\catcode`@=12

\title{LIGHT-QUARK, HEAVY-QUARK SYSTEMS: AN UPDATE}

\author{BENJAMIN GRINSTEIN\\
        \em SSC Laboratory, 2550 Beckleymeade Ave., MS-2007\\
        \em Dallas, Texas 75248 USA}

\begin{document}
\vspace*{1.0in}

\hfill SSCL-Preprint-514\break
\hfill hep-ph/9310362 \break

\vspace{1.0in}

{\bf
\centerline{SOME APPLICATIONS OF HEAVY QUARK EFFECTIVE THEORY}
\vspace{1.0in}
\centerline{BENJAMIN GRINSTEIN}
\vspace{.3cm}
\centerline{{\em SSC Laboratory (Defunct), 2550 Beckleymeade Ave., MS-2007}}
\centerline{{\em Dallas, Texas 75248, USA}}
}
\vspace{1.0in}
To appear in ``The Proceedings of
 the Workshop on b Physics at Hadron Accelerators''
\newpage

\maketitle
 \pagestyle{empty}  

\begin{quote}
We review many of the recently developed applications of Heavy Quark
Effective  Theory techniques. After a brief update on Luke's theorem,
we describe striking relations between heavy baryon form factors, and
how to use them to estimate the accuracy of the extraction of
$|V_{cb}|$. We discuss factorization and compare with experiment. An
elementary presentation, with sample applications, of
reparametrization invariance comes next. The final and most extensive
chapter in this review deals with phenomenological lagrangians that
incorporate heavy-quark spin-flavor as well as light quark chiral
symmetries. We compile many interesting results and discuss the
validity of the calculations.
\end{quote}

\onehead{1.}{INTRODUCTION}
It seems hardly appropriate to devote any time to reviewing the
fundamentals of Heavy Quark Effective Theory (HQET), both because this
is a meeting of experts and because several good reviews of the
subject are now available.\cite{MBWise,Bgreview} Instead
of wasting any space
introducing conventions, I simply choose to use the notation of
Ref.~2. Thus, I will be able to devote more energy towards a
description of recent developments in this field.

I view this paper as updating and expanding on Ref.~2. There the HQET
was presented and a few applications discussed at length. Other
applications where briefly discussed. Much has changed since Ref.~2 was
written, and it seems the time is ripe for an extension of that work.
Because of time and space limitations this is not intended as an
extensive overview of progress in the field since Ref.~2 was written.
Rather, I shall pick and choose according to my taste, familiarity
with the subjects, and what I perceived as relevant to the
participants of the workshop.

\onehead{2.}{AN UPDATE ON LUKE'S THEOREM}

Presumably the best known consequence of heavy quark symmetries is
that the form factors for semileptonic $B\to D$ and $B\to D^*$ decays
are determined at the point of zero recoil (equal $B$ and $D$
velocities). Luke's theorem states that this normalization of the
meson form factors has no $1/M_Q$ corrections.\cite{luke} It is not
widely appreciated that Luke's original proof did not exclude possible
{\sl short distance\/} corrections of order $(\bas(m_c)/m_c)$. It
turns out it is easy to extend Luke's proof to exclude corrections of
this sort to any order in the strong coupling.\cite{choben}

Similarly, the normalization of form factors for
$\Lambda_b\to\Lambda_c$ semileptonic decay is computable up to
corrections of order $1/M_Q^2$.\cite{choben,ggw}

\onehead{3.}{HEAVY BARYON FORM FACTOR RELATIONS}

\ttwohead{3.1}{Relations to First Order in $1/M_Q$} 

Six form factors encode the semileptonic decay amplitude $\Lambda_b
\to\Lambda_c e\overline{\nu}$. The transition lends itself
particularly well to HQET analysis because it is tightly constrained
by the heavy quark spin symmetry.\cite{baryonrefs} Like their mesonic
counterparts, the six form factors that parameterize this baryonic
process are predicted at leading order in the $1/M_Q$ expansion in
terms of a single Isgur-Wise function.  In contrast with their mesonic
counterparts, one can prove that this is still the case at order
$1/M_Q$.\cite{ggw} In other words, five relations among these six form factors
remain after $O(1/m_c)$ and $O(1/m_b)$ corrections are included.

Remarkably, that such relations can be written is not precluded by short
distance effects to any order in the strong coupling constant.\cite{choben}
However the relations themselves get corrected order by order in
perturbation theory. To see how this works, define the form factors
through
\begin{eqnarray}
\langle\Lambdac|V^\mu|\Lambdab\rangle &=& \ubar[F_1(\vv)
\gamma^\mu + F_2(\vv) v^\mu   + F_3(\vv) v'^\mu ]\uvs \label{eq:ffsv}  \\
\vev{\Lambdac|A^\mu|\Lambdab} &=& \ubar [G_1(\vv) \gamma^\mu
+ G_2(\vv) v^\mu
  + G_3(\vv) v'^\mu ]\gamma^5 \uvs 		\label{eq:ffsa}
\end{eqnarray}
where $v$ and $s$ refer to the velocity and spin of the state
$\Lambda_b$ and of the Dirac spinor $u$. Then, the relations between
form factors are\cite{choben}
\begin{eqnarray}
{F_1 \over G_1}& =& 1+\Bigl[ \cLambdabar + \bLambdabar \Bigr]  {2
\over (\vv+1)}  +
\fourthirds\coupling r
  + \fourthirds\coupling\cLambdabar
  {2(1+r-\vv r) \over (\vv+1) } \label{eq:relone}\\
{F_2 \over G_1}& = &{G_2 \over G_1} = - \cLambdabar {2 \over (\vv+1) }
- \fourthirds \coupling r - \fourthirds\coupling \cLambdabar
{2(1+r-\vv r) \over (\vv+1) } \label{eq:reltwo}\\
{F_3 \over G_1} &=& -{G_3 \over G_1} =
 - \bLambdabar {2 \over (\vv+1) } \label{eq:relthree}
\end{eqnarray}
where
\eqn{
r = {\log(\vv +\sqrt{(\vv)^2-1}) \over \sqrt{(\vv)^2-1}}~.
}
and ${\overline \Lambda} $ is an undetermined constant with unit mass
dimensions, expected to be of order of the hadronic scale, ${\overline
\Lambda} \sim500$~MeV. If in Eqs.~\ref{eq:relone} -- \ref{eq:relthree}
one sets $\bas(m_c)=0$ and $\overline\Lambda=0$, one recovers the
zeroth order results of Ref.~6, 
while the results of Ref.~5 
are obtained by allowing $\overline\Lambda\ne0$ but with
$\bas(m_c)=0$. Clearly there are also corrections of order $\bas(m_b)$
and of higher order in $1/M_Q$.

Heavy quark symmetries give the value of the form factors at zero
recoil. In the leading-log approximation
\eqn{
G_1(1)=\left({\bas(m_b)\over\bas(m_c)}\right)^{a_I} \label{eq:gone}
}
There are no corrections of order $1/M_Q$ to this
relation.\cite{ggw,choben} The counterpart of this prediction for
mesons is used in the measurement of the mixing angle $|V_{cb}|$.

The form factor relations \ref{eq:relone} -- \ref{eq:relthree} provide
a valuable means for assessing the uncertainty in future measurements
of the mixing angle $|V_{cb}|$.  It is reasonable to expect the
prediction in Eq.~\ref{eq:gone} to hold to the same accuracy with
which the form factors satisfy the predicted relations, at least for
small or moderate $\vv-1$.

\twohead{3.2}{Relations To All Orders In $1/m_c$}

The relations above were obtained by expanding both in $1/m_c$ and
$1/m_b$. Because the charm quark is only a few times heavier than
typical hadronic scales, the corrections to the relations
\ref{eq:relone} -- \ref{eq:relthree} may be large. Remarkably,
Mannel and Roberts obtain four relations among the six form factors
without assumptions on the size of $m_c$.\cite{mannrob} Expanding in
$1/m_b$, \ie using the HQET for the $b$ quark, the spin symmetry
acting on the $b$ quark alone is enough to limit to two the number of
independent form factors in $\Lambda_b\to\Lambda_q$, where $q= u,c$:
\eqn{
\vev{\Lambda_q(p',s')|\bar q\Gamma h^{(b)}_v|\Lambdab} =
 {\overline u}(p',s')[f_1(v\ccdot p')+\vslash f_2(v\ccdot
p')]\Gamma\uvs                      \label{eq:lightc}
}
It is straightforward to write the six form factors in
Eqs.~\ref{eq:ffsv} -- \ref{eq:ffsa} in terms of the two form factors
in Eq.~\ref{eq:lightc}. Explicit relations between the form factors
follow from eliminating $f_{1,2}$ from Eq.~\ref{eq:lightc}:
\eqna{
F_1 &=& G_1  - G_2 \label{eq:mrone}\\
F_2 &=& G_2 \\
F_3 &=& 0 \\
G_3 &=& 0 \label{eq:mrfour}
}
These remarkably simple expressions receive corrections in order $1/m_b$ and
$\bas(m_b)/\pi$, but are valid for arbitrary $m_q$ (provided
$m_q<m_b$). Moreover, the perturbative corrections $\sim\bas(m_b)/\pi$ are
computable; the leading correction is obtained by replacing\cite{fggw}
\eqn{
\Gamma \to \Gamma - {\bas(m_b)\over 6\pi}\g_\mu\vslash\Gamma\vslash \g^\mu
}
in Eq.~\ref{eq:lightc}.

By taking the limit $m_b\to\infty$, one readily checks
that  Eqs.~\ref{eq:relone} -- \ref{eq:relthree} are consistent with
Eqs.~\ref{eq:mrone} -- \ref{eq:mrfour}.

\onehead{4.}{FACTORIZATION}

\ttwohead{4.1}{Summary of Theory}
Consider purely hadronic $B$-meson decays into singly charmed final
states. I have in mind the class of processes that includes $B\to
D\pi$, $B\to D^*\pi$, $B\to D\rho$, etc. The interaction Hamiltonian
density mediating these decays is
\eqn{
{\cal H} = \frac{G_F}{\sqrt2}V_{cb}V^*_{ud} [c_1 \bar b_L \g_\mu c_L
\bar u_L \g^\mu d_L + c_2 \bar b_L \g_\mu T^a c_L \bar u_L \g^\mu T^a
d_L]~,\label{eq:hamilton}
}
where $c_{1,2}$ are calculable short distance QCD corrections, $T^a$
are color octet matrices, and $q_L$ stands for a left handed
quark. The second term in ${\cal H}$ arises from short distance QCD
effects. Factorization in a particular decay, say $B\to D\pi$ is the
statement that the following equation is true:
\eqn{
\vev{D\pi | {\cal H} | B} = \frac{G_F}{\sqrt2}V_{cb}V^*_{ud}
    c_1 \vev{D | \bar b_L \g_\mu c_L | B}
	\vev{\pi | \bar u_L \g^\mu d_L |0}      \label{eq:factorize}
}
If factorization holds, the rate for the hadronic decay (the left hand
side in eq.~(9)) is given in terms of a meson decay constant
($\vev{\pi(q) | \bar u_L \g^\mu d_L |0}= i f_\pi q^\mu $) and the form
factors for $B\to D$ at a fixed momentum transfer (that is $\vev{D |
\bar b_L \g_\mu c_L | B}$ at $q^2=M_\pi^2$).

Whether a particular matrix element factorizes is a dynamical issue
that involves non-perturbative strong interactions, and is therefore
hard to settle from first principles. We do know, nevertheless, that
factorization does not hold for a large class of two body decays. In
the case of $K$ decays, the $\Delta I=1/2$ rule is a stark reminder
that simple factorization does not hold. More recently, a wealth of
evidence {\sl against\/} factorization in $D$-meson decays (as in
$D\to K\pi$) has been amassed.\cite{yamamoto}

To my knowledge there are two known theoretical approaches to
demonstrating factorization. It holds in leading order in the $1/N_c$
expansion, where $N_c$ is the number of colors in
QCD.\cite{largenfact} And it holds in the leading order in the $1/M_Q$
expansion.\cite{duganben}

Now, these approaches are rather different. The large $N_c$ limit is
fairly democratic: effectively, it predicts factorization in any meson
decay into two meson final states, regardless of which flavors are
involved in the transition. It does not predict, as far as I can tell,
factorization in baryon decays (because the number of non-spectator diagrams,
each suppressed by $1/N_c$, scales like $N_c$).

The large $M_Q$ limit is fairly restrictive as to which transitions may
exhibit factorization. It must be a transition of the form
${\rm M}\to{\rm M}' {\rm X}$ where M and M$'$ are heavy hadrons, with their
masses in a fixed ratio, both scaling with the large parameter $M_Q$,
and X is a hadronic state with small invariant mass, that is, it's
mass does not grow with $M_Q$. To the extent that the $b$ and $c$
quarks can be considered  heavy, this approach can be used for $B\to
D\pi$, and even for baryons as in $\Lambda_b\to\Lambda_c\pi$. But in
the case of $D$ decays this approach says nothing, since the final
state does not involve any heavy quarks.

I will have nothing to say about phenomenological approaches to
factorization.\cite{BS} My interest here is on what can be obtained
from first principles, even if only in some approximation. Clearly we
have a better chance of learning about dynamics if we concentrate on
results that follow directly from QCD than on phenomenological
approaches. It is for this reason also that we have nothing to say
about decays such as $B\to\psi K$ which may very well factorize, but
we don't know of any first principles justification for that to be the
case. (In fact, one expects factorization in  the inclusive resonant rate
$B\to\psi X_s$, where by resonant we mean that the $\psi$ is directly
produced. P-wave  charmonium production in $B$-meson decays is known
not to factorize.\cite{bodwin} Consequently nonresonant inclusive $\psi$
production won't either).

\twohead{4.2}{Comparison With Experiment}
The large $N_c$ approach is far too democratic: experimentally it is
found that factorization does not hold in decays of heavy mesons to
light mesons, or in light-to-light decays. In this section I intend to
investigate the predictions of the large mass limit as far as
factorization is concerned.

We start by considering qualitative statements implied by the
arguments of Ref.~11. Feynman diagrams that don't factorize on account
of the light quark in the initial heavy meson ending up in the light
hadron in the final state are suppressed by $1/M_Q$. Now,
the only diagrams that contribute to $\bar B^0 \to D^0\pi^0$ are of
this kind (and therefore $\bar B^0 \to D^0\pi^0$ does not itself
factorize). Hence if factorization is to hold to some accuracy
$\epsilon$, the rate for $\bar B^0 \to D^0\pi^0$ ought to be
suppressed relative to the rate for $\bar B^0 \to D^+\pi^-$ or $ B^-
\to D^0\pi^-$ by roughly $\epsilon^2$.

A quick glance at the particle data book shows that $\bar B^0$ decays
into $D^+\pi^-$, $D^+\rho^-$, $D^+a_1(1260)^-$, $D^*(2010)^+\pi^-$,
$D^*(2010)^+\rho^-$ and $D^*(2010)^+a_1(1260)^-$ have been observed
and have branching fractions in the 0.3\% to 1.8\% range. Non of the
corresponding decays into $D^0$ or $D^*(2010)^0$ plus a neutral light
meson have been observed. An upper bound exists on the branching
fraction for $\bar B^0 \to D^0 \rho^0$ of $6\times10^{-4}$. This is
all as expected from the factorization argument in the paragraph
above.

Quantitative, model independent,\cite{reader} tests of factorization
are readily available. We will consider three kinds of such tests. The
first two compare different two body decays which are related by a
combined use of factorization and either isospin or heavy quark spin
symmetries. In the third we compare some two body decays to
corresponding semileptonic rates. The third is the most direct test,
but is not available for as many processes. Also, it is interesting to
see how well the other symmetries, and in particular heavy quark spin
symmetry, work.

Using isospin symmetry on the factorized amplitudes, one obtains that
the partial widths for the charged and the neutral meson decays into
charmed two body decays should be equal.  That is, one expects
$\Gamma(\bar B^0 \to D^+\pi^-)\approx \Gamma(B^-
\to D^0\pi^-)$ and similar relations for the other modes. These results
are not predicted by isospin symmetry alone. The hamiltonian in
Eq.~\ref{eq:hamilton} has $\Delta I=0,1$, while the $B$ and $D$ mesons
are both $I=1/2$ states, so the final $D\pi$ state is a combination
of $I=1/2$ and $I=3/2$. There are three independent amplitudes, but
they are not independent if factorization holds.

This can be tested assuming the total widths of the charged and
neutral $B$-mesons are equal. It is seen that these relations
hold to the present experimental accuracy. For example, the particle
data book gives
\eqn{
{\rm Br}(B^-\to D^0\pi^-) = (3.8\pm1.1)\times10^{-3}
}
while
\eqn{
{\rm Br}(\bar B^0 \to D^+\pi^-) = (3.2\pm0.7)\times10^{-3} \label{eq:btodpim}
}
and similar results for the other three modes mentioned above.

Since the factorized amplitude is given in terms of the semileptonic
form factors, one can use heavy quark spin symmetry to relate the
rates into $D$ and $D^*$ final states:
\eqn{
\Gamma(\bar B\to D X) = \Gamma(\bar B\to D^* X)~.
}
This seems to work well, too. For example, from the particle data book
\eqn{
{\rm Br}(\bar B^0 \to D^*(2010)^+\pi^-) = (3.2\pm0.7)\times10^{-3} }
to be compared with ${\rm Br}(\bar B^0 \to D^+\pi^-)$ in
Eq.~\ref{eq:btodpim} above. It is remarkable that both factorization
and heavy quark spin symmetry can be tested simultaneously and that
both seem to work rather well.

Table 1 shows CLEO II measured branching fractions.\cite{somecleoii}
The two columns are related by spin symmetry (if factorization
holds). We group lines into pairs for the neutral and charged $B$
decays. Thus the combined result of factorization, isospin symmetry,
heavy quark spin symmetry and the assumption of equal $B^0$ and $B^+$
lifetimes, is that all entries in each $2\times2$ block are equal. It
can be seen that, within experimental errors this is the case. It is
intriguing that the central values of all of the $\bar B^0$ decays are
about 70\% of the corresponding $B^-$. If this is a real effect it
could be evidence against factorization. It could also be interpreted
as evidence for different $B^0$ and $B^+$ lifetimes,
$\tau(B^0)/\tau(B^+)\sim0.7$. But this is hard to reconcile with
direct results from the DELPHI\cite{delphiratio} and
ALEPH\cite{alephratio} experiments, which tend to favor
$\tau(B^0)/\tau(B^+)>1$.

\vspace{0.75cm}

\vbox{
\centerline{Table~1. Some CLEO II Branching Fractions}
\begin{center}
\begin{tabular}{|c|c||c|c|}
\hline
Decay & Branching & Decay & Branching  \\
    & Fraction & & Fraction \\
\hline
$B^-\to D^0\pi^-$ & $0.40\pm0.03\pm0.09 $
& $B^-\to D^*(2010)^0\pi^-$ & $0.35\pm0.05\pm0.12 $ \\
$\bar B^0 \to D^+\pi^-$ & $0.26\pm0.03\pm0.06$
& $\bar B^0 \to D^*(2010)^+\pi^-$ & $ 0.27\pm0.04\pm 0.06$ \\
\hline
$B^-\to D^0\rho^-$ & $1.02\pm0.11\pm0.29 $
& $B^-\to D^*(2010)^0\rho^-$ & $1.14\pm0.16\pm0.37 $ \\
$\bar B^0 \to D^+\rho^-$ & $0.71\pm0.10\pm0.21$
& $\bar B^0 \to D^*(2010)^+\rho^-$ & $ 0.73\pm0.10\pm 0.16$ \\
\hline
\end{tabular}
\end{center}
}
\vspace{0.7cm}

If factorization holds, the degree of polarization in the decay $\bar
B^0 \to D^*(2010)^+\rho^-$ can be predicted in terms of the degree of
polarization in the semileptonic decay:\cite{somecleoii}
\eqn{
{\Gamma_L\over\Gamma}(\bar
B^0 \to D^*(2010)^+\rho^-) =
{d\Gamma_L\over d\Gamma}(\bar
B^0 \to D^*(2010)^+\ell\nu)|_{m^2_{\ell\nu}=m^2_\rho}
}
Here the differential rates on the right hand side are with respect to
the invariant lepton pair mass, $m^2_{\ell\nu}$. The CLEO
collaboration finds
\eqn{
{\Gamma_L\over\Gamma}(\bar
B^0 \to D^*(2010)^+\rho^-) = 0.90 \pm 0.07 \pm 0.05
}
while the expected value from the semileptonic decay is 85\% -- 88\%.

Finally, the most direct test of factorization is obtained by
comparing directly both sides of Eq.~\ref{eq:factorize}, or
equivalently by testing whether Bjorken's ratio
\eqn{
R_\pi\equiv{\Gamma(\bar B^0 \to D^*(2010)^+\pi^-) \over
d\Gamma(\bar B^0 \to D^*(2010)^+\ell\nu )/dm^2_{\ell\nu}
|_{m^2_{\ell\nu}=M^2_\pi}}
}
agrees with the expectation from factorization:
\eqn{
R_\pi=6\pi^2f_\pi^2 c_1^2 }
Similar expressions can be written with the pion replaced by some
other final state.  Experimentally, the ratios $R_\pi$ and $R_\rho$
for the neutral meson decay have been studied. The results of CLEO~II
measurements and the expectations from factorization are summarized
in Table~2.\cite{somecleoii}
\vspace{0.75cm}

\vbox{
\centerline{Table~2. CLEO II Results on Bjorken's Ratios}
\begin{center}
\begin{tabular}{|c|c|c|}
\hline
  & Experiment & Factorization \\
\hline
$R_\pi$& $1.3\pm0.2\pm0.3$ & $1.2\pm0.2$ \\
$R_\rho$ & $3.2\pm0.4\pm0.7$ & $3.3\pm 0.6$ \\
\hline
\end{tabular}
\end{center}
}


\onehead{5.}{REPARAMETRIZATION INVARIANCE}
There is an ambiguity in assigning a four-velocity, $v$, and residual
momentum, $k$, to a particle in the HQET. Recall that only the
momentum $p=Mv+k$ has physical significance. One may shift both the
velocity and residual momentum to obtain the same physical
momentum:
\eqna{
v &\to& v +q/M\\ k &\to& k-q
}
The only constraint on the vector q is that the new four-velocity be
properly normalized:
\eqn{
(v +q/M)^2 = 1
}
The effective field theory must be invariant under these
reparametrizations.\cite{dgg} The reparametrizations mix
different orders in $1/M$. Hence, one can use reparametrization
invariance to put constraints on the form of the $1/M$
corrections.\cite{manluk}

As an example of an application consider the matrix element of the
vector current between two pseudoscalar mesons. When using the HQET to order
$1/M$ it is important to include in the description of the states both
the velocity label $v$ and the residual momentum $k$:
\eqn{
\vev{v,k' | V_\mu | v,k} = f_1 v_\mu + f_2 (k_\mu+k'_\mu) +
 f_3(k_\mu-k'_\mu)~.
}
Here $V_\mu$ stands for the heavy quark current including $1/M$
corrections. Now, in the ``full theory'', that is, the theory without
any large mass expansion, there are only two independent form factors,
usually denoted by $f_+$ and $f_-$. It shouldn't be necessary to
introduce three form factors in the effective theory. This is implied
by reparametrization invariance, which gives the relation
\eqn{
f_2 = {1\over2M}f_1}

Of more practical importance is the use of reparametrization
invariance to constrain the form of the heavy quark current in the
effective theory. The heavy quark vector current has a $1/M$
expansion\cite{Bgreview}
\eqn{
\sum_iC^{(i)}(\vv) \CO^{(0)}_i + {1\over2M_Q}\sum_j D^{(j)}(\vv) \CO^{(1)}_j
+ {1\over2M_{Q'}}\sum_j D^{\prime(j)}(\vv) \CO^{(1)}_j \label{eq:vectcur}
}
where $\CO^{(0)}_i$ and $\CO^{(1)}_j $ stand for vector operators of
dimension three and four respectively with $\overline Q'_{v'} Q_{v}$
quantum numbers, and their coefficients $C$, $D$ and $D'$ are
perturbatively calculable. For example, at tree level the current is
\eqn{
 \bqvqv{\g_\mu} +
{1\over2M_Q}\bqvqv{\g_\mu i\overslash{D}}-
{1\over2M_{Q'}}\bqvqv{i\larr\overslash{D}\g_\mu}
 }
where we have used the equations of motion, $v\ccdot D Q_v=0$.
Now, the vector current in Eq.~\ref{eq:vectcur} will be
reparametrization invariant if and only if it depends on the
velocities $v$ and $v'$ in the combinations
\eqn{
v_\mu + k_\mu/M_Q \qquad\hbox{ and}\qquad v'_\mu + k'_\mu/M_{Q'}
}
or in operator language
\eqn{
v_\mu + iD_\mu/M_Q \qquad\hbox{ and}\qquad v'_\mu - i\larr D_\mu/M_{Q'}
}

Consider, for example, the following leading term in Eq.~\ref{eq:vectcur}
\eqn{
C^{(1)}(\vv)\bqvqv{\g_\mu}=
\bqvqv{\opvp C^{(1)}(\vv)\g_\mu\opv}
}
It must appear in the following combination to be invariant under
separate reparametrizations of $v$ and $v'$
\eqna{
\bqvqv{{\left({1+\vslash'-i\larr\overslash{D}/M_{Q'}\over2}\right)}
C^{(1)}((v'-i\larr D/M_{Q'})\cdot (v+iD/M_Q))\g_\mu
{\left({1+\vslash +i D/M_Q\over2}\right)}} \nonumber\\
=C^{(1)}(\vv)\left[\bqvqv{\g_\mu} +
{1\over2M_Q}\bqvqv{\g_\mu i\overslash{D}}-
{1\over2M_{Q'}}\bqvqv{i\larr\overslash{D} \g_\mu} \right] \nonumber\\
+ {dC^{(1)}\over d \vv}\left[{1\over M_Q}\bqvqv{\g_\mu v'\ccdot D} -
 {1\over M_{Q'}}\bqvqv{v \ccdot \larr D\g_\mu }\right] + \cdots
}
In a similar manner the coefficients of other dimension four operators
can be constrained by applying the same method to the other two
dimension three operators, $\bqvqv{v_\mu}$ and $\bqvqv{v'_\mu}$.

The calculation leading to the $1/m_c$ corrections in
$\Lambda_b\to\Lambda_ce\nu$ required the coefficients of the vector and
axial currents to order $1/m_c$. It is easy to check that the
coefficients used\cite{choben} to obtain the relations in
Eqs.~\ref{eq:relone} -- \ref{eq:relthree} satisfy the constraints from
reparametrization invariance. The calculation there would have been
simplified vastly had reparametrization invariance been used to obtain
the result. (Alternatively, reparametrization invariance gives an
independent test of the calculation).

\onehead{6.}{CHIRAL SYMMETRY TOO}
\ttwohead{6.1}{Generalities}
Chiral symmetry and soft pion theorems have been used in particle
physics for several decades now with great success. The most efficient
way of extracting information from chiral symmetry is by writing a
phenomenological lagrangian for pions that incorporates both the
explicitly realized vector symmetry and the non-linearly realized
spontaneously broken axial symmetry.\cite{georgi} Theorems that
simultaneously use heavy quark symmetries and chiral symmetries are
most expediently written by means of a phenomenological lagrangian for
pions and heavy mesons that incorporates these
symmetries.\cite{chiral,yanetal}

In the limit $m_b\to\infty$, the $\ol B$ and the $\ol B^*$ mesons are
degenerate, and to implement the heavy quark symmetries it is
convenient to assemble them into a ``superfield'' $H_a(v)$:
\eqn{
H_a(v)={1+\vslash\over2}\left[
    \ol B_a^{*\mu}\g_\mu-\ol B_a\g^5\right]\,.\label{eq:superf}}
Here $v^\mu$ is the fixed four-velocity of the heavy meson, and $a$ is
a flavor $SU(3)$ index corresponding to the light antiquark. Because
we have absorbed mass factors $\sqrt{2m_B}$ into the fields, they have
dimension 3/2; to recover the correct relativistic normalization, we
will multiply amplitudes by $\sqrt{2m_B}$ for each external $\ol B$ or
$\ol B^*$ meson.

The chiral lagrangian contains both heavy meson superfields and
pseudogoldstone bosons, coupled together in an $SU(3)_L\times SU(3)_R$
invariant way.  The matrix of pseudogoldstone bosons appears in the
usual exponentiated form $\xi=\exp(\im{\cal M}/f)$, where
\eqn{
    {\cal M}=\pmatrix{{\textstyle{1\over\sqrt{2}}}\pi^0+{\textstyle
    {1\over\sqrt{6}}}\eta&\pi^+&K^+\cr\pi^-
    &{\textstyle{-{1\over\sqrt{2}}}}\pi_0+
    {\textstyle{1\over\sqrt{6}}}\eta&K^0\cr K^-&\overline K^0&-
    {\textstyle\sqrt{2\over 3}}\,\eta}\,,}
and $f$ is the pion (or kaon) decay constant.  The bosons couple to
the heavy fields through the covariant derivative and axial vector
field,
\eqna{
    D_{ab}^\mu&=&\delta_{ab}\del^\mu+V_{ab}^\mu
    =\delta_{ab}\del^\mu+\half\left(\xi^\dagger
    \del^\mu\xi+\xi\del^\mu\xi^\dagger\right)_{ab}\,, \\
    A_{ab}^\mu&=&{\im\over 2}\left(\xi^\dagger\del^\mu\xi
    -\xi\del^\mu\xi^\dagger\right)_{ab}
    =-{1\over f}\partial_\mu{\cal M}_{ab}+{\cal O}({\cal M}^3)\,.}
Lower case roman indices correspond to flavor $SU(3)$. Under chiral
$SU(3)_L\times SU(3)_R$, the pseudogoldstone bosons and heavy meson
fields transform as $\xi\to L\xi U^\dagger=U\xi R^\dagger$, $A^\mu\to
UA^\mu U^\dagger$, $H\to HU^\dagger$ and $(D^\mu H)\to (D^\mu
H)U^\dagger$, where the matrix $U_{ab}$ is a nonlinear function of the
pseudogoldstone boson matrix ${\cal M}$.

The chiral lagrangian is an expansion in derivatives and pion fields,
as well as in inverse powers of the heavy quark mass.  The kinetic
energy terms take the form
\eqn{
    {\cal L}_{\rm kin}={1\over8}
    f^2\,\partial^\mu\Sigma_{ab}\,\partial_\mu \Sigma^{\dagger}_{ba}
    -\Tr\left[\overline H_a(v)i v\cdot D_{ba} H_b(v)\right]\,,}
where $\Sigma=\xi^2$. Here the trace is in the space of $4\times 4$
Dirac matrices that define the ``superfields'' $H_a(v)$ in
Eq.~\ref{eq:superf}.  The leading interaction term is of dimension
four,
\eqn{
	{\cal L}_{\rm int}=
    g\,\Tr\left[\ol H_a(v)H_b(v)\,\Aslash_{ba}\g^5\right]\,,
\label{eq:gterm}}
where $g$ is an unknown parameter, of order one in the constituent
quark model. The analogous term in the charm system is responsible for
the decay $D^*\to D\pi$. Expanding the term in the lagrangian
in~\ref{eq:gterm} to linear order in the Goldstone Boson fields,
${\cal M}$, we find the explicit forms for the $D^*D{\cal M}$ and
$D^*D^*{\cal M}$ couplings
\eqn{
\left[\left({-2g\over f}\right)D^{*\nu}\del_\mu{\cal M} D^\dagger +
{\rm h.c.}\right] + \left({2gi\over f}\right)\emnlk
D^{*\mu}\del^\nu{\cal M} D^{*\lambda} v^\kappa\,.\label{eq:gexpand}
}
Using this one can compute the partial width
\eqna{
\Gamma(D^{*+}\to D^0\pi^+) &=& {g^2\over6\pi f^2}|\vec p_\pi|^3 \\
\Gamma(D^{*+}\to D^+\pi^0) &=& {g^2\over12\pi f^2}|\vec p_\pi|^3
}
The ACCMOR collaboration has reported an upper limit of 131~KeV on the
$D^*$ width.\cite{accmor} The branching fractions for $D^{*+}\to
D^0\pi^+$ and $D^{*+}\to D^+\pi^0$ are $(68.1\pm1.0\pm1.3)$\% and
$(30.8\pm0.4\pm0.8)$\%, respectively, as measured by the CLEO
collaboration.\cite{cleobfs} Using $f=130$~MeV, one obtains the
limit $g^2<0.5$. Even if the $D^*$ decay width is too small to
measure, radiative $D^*$ decays provide an indirect means for
determining the coupling $g$, and provide a lower bound
$g^2\gs0.1$.\cite{jimcho}

Since charmed and beauty baryons are long lived, one can write down
phenomenological lagrangians for their interactions with pions. These
are as well justified and should be as good an approximation as the
lagrangian for heavy mesons discussed above. The treatment is rather
similar, and due to space limitations, we refer the interested reader
to the literature.\cite{yanetalbaryons}

\twohead{6.2}{$B\to De\nu$ and $B\to D^*\pi e \nu$}
As a first example of an application consider a soft pion theorem
that relates the amplitudes for $B\to D^*e \nu$ and $B\to D^*\pi e
\nu$.\cite{yanetal} The heavy quark current is represented in the
phenomenological lagrangian approach by
\eqn{
J^{\bar c b}_\mu = \bar h^{(c)}_{v'} \g_\mu(1-\g_5)h^{(b)}_v \to
\xi(\vv)\Tr\overline H_a^{(c)}(v')\g_\mu(1-\g_5) H_a^{(b)}(v)
+\cdots \label{eq:effcurr}
}
where the ellipsis denote terms with derivatives, factors of light
quark masses $m_q$, or factors of $1/M_Q$, and $\xi(\vv)$ is the
Isgur-Wise function. The leading term in Eq.~\ref{eq:effcurr} is
independent of the pion field. Therefore, it is pole diagrams that
dominate the amplitude for semileptonic $B\to D\pi$ and $B\to D^*\pi$
transitions; see Fig.~1. These pole diagrams are calculable in this
approach, and are determined by the Isgur-Wise function and the
coupling $g$.

\vskip1.5truecm
 \epsfbox{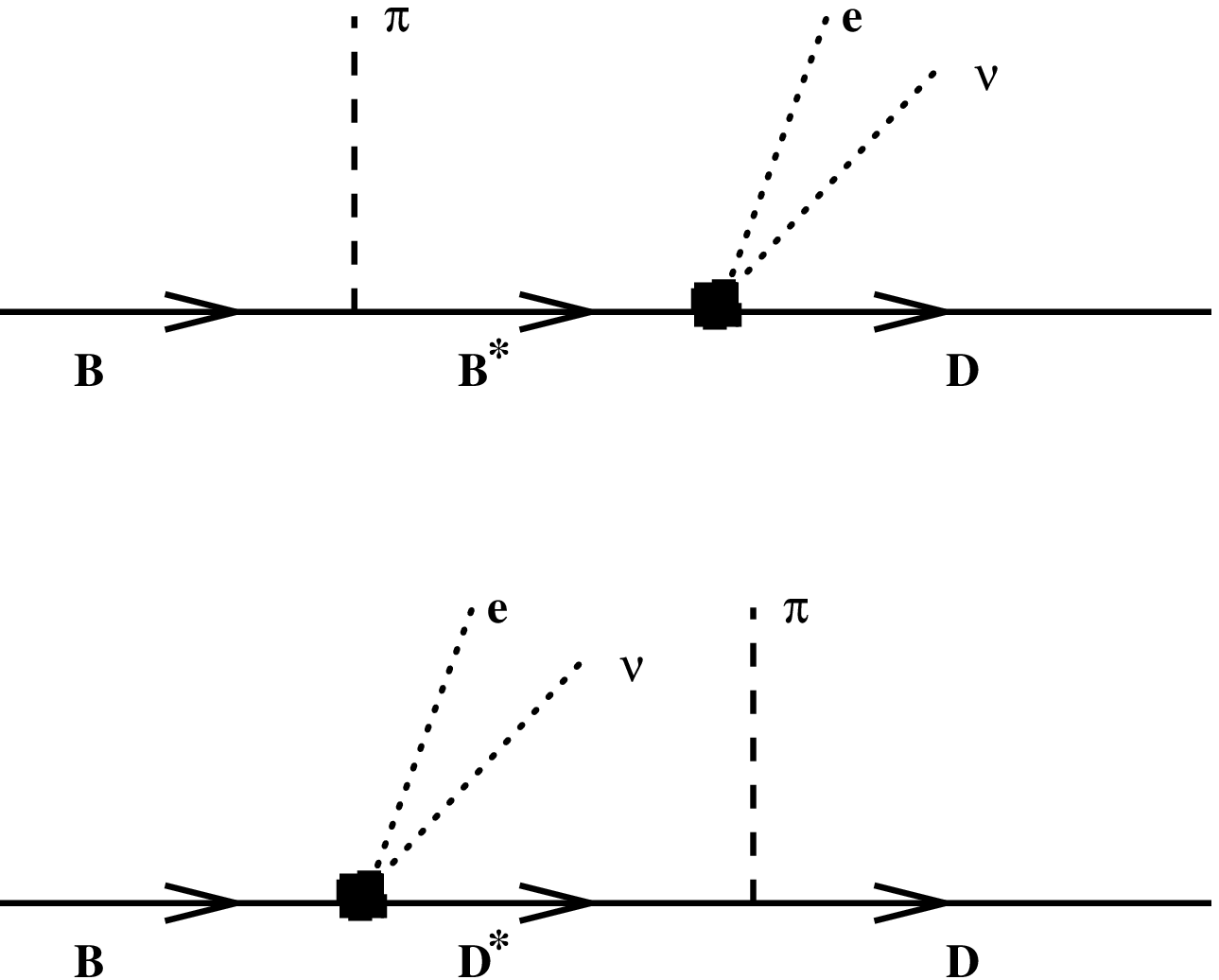}
\centerline{Figure 1. Feynman diagrams for $B\to De\nu$}
\vskip2.0truecm
A straightforward calculation gives
\eqna{
\vev{D(v')\pi^a(q)|J^{\bar c b}_\mu|B(v)} &=&
iu(B)^*\half\tau^au(D)\sqrt{M_BM_D}{g \over f} \xi(\vv) \nonumber\\
&\times& \left\{{1\over v\ccdot q}[i\emnlk q^\nu
v^{\prime\lambda}v^\kappa + q\cdot(v+v')v_\mu - (1+\vv)q_\mu
                                                \right. \nonumber\\
&{}&\left. -{1\over v'\ccdot q}[i\emnlk q^\nu
v^{\prime\lambda}v^\kappa + q\cdot(v+v')v'_\mu - (1+\vv)q_\mu \right\}
}
where $u(M)$ stands for the isospin wavefunction of meson $M$. A
similar but lengthier expression is found for $B\to D^*\pi e
\nu$.\cite{yanetal} If the coupling $g$ is close to its upper limit,
this process could be an important correction to the inclusive
semileptonic rate. It may, perhaps, account for some of the
anomalously large ``$D^{**}$'' contributions observed by
CLEO.\cite{morecleo}

\twohead{6.3}{Violations To Chiral Symmetry}
Phenomenological lagrangians are particularly well suited to explore
deviations from symmetry predictions. In the context of heavy mesons,
several quantities of considerable interest have been studied.
Moreover, the self-consistency of the approach has been explored. It
would be impossible to cover all of this in this talk. I will briefly
comment on a few of those results, and invite you to consult the
references for further details.

In order to study violations of chiral symmetry, one must introduce
symmetry breaking terms into the phenomenological lagrangian. The
light quark mass matrix $m_q={\rm diag}(m_u, m_d, m_s)$ parametrizes
the violations to flavor $SU(3)_V$. To linear order in $m_q$ and
lowest order in the derivative expansion, the correction to the
phenomenological lagrangian is
\eqna{
\Delta{\cal L} &=&
\lambda_0\ \left[ m_q \Sigma + m_q \Sigma^\dagger \right]^a{}_a
				\nonumber\\
&+&\lambda_1 \Tr \bar H^{(Q)a} H_b^{(Q)} \left[ \xi m_q \xi + \xi^\dagger
m_q \xi^\dagger \right]^b{}_a		\nonumber\\
&+& \lambda_1^\prime \Tr \bar H^{(Q)a} H_a^{(Q)}\
\left[ m_q \Sigma + m_q \Sigma^\dagger \right]^b{}_b \label{eq:lcoup}}
The coefficients $\lambda_0$, $\lambda_1$ and $\lambda'_1$ are
determined by non-perturbative strong interaction effects, but may be
determined phenomenologically. We postpone consideration of mass
relations obtained from this lagrangian until we have introduced
heavy quark spin symmetry breaking terms into the lagrangian too.

The decay constants for the $D$ and $D_s$ mesons, defined by
\eqn{
\vev{ 0 | \bar d \gamma_{\mu} \gamma_5 c| D^+(p)}
= i f_D p_{\mu}
}
and
\eqn{
\vev{ 0 | \bar s \gamma_{\mu} \gamma_5 c | D_s(p)}
= i f_{D_s} p_{\mu}~,
}
determine the rate for the purely leptonic decays $D^+ \rightarrow
\mu^+ \nu_{\mu}$ and $D_s \rightarrow \mu^+ \nu_{\mu}$. These are
likely to be measured in the future.\cite{cleofds} In the chiral
limit, where the up, down and strange quark masses go to zero, flavor
$SU(3)_V$ is an exact symmetry and so $f_{D_S} / f_D = 1$.  However
$m_s \neq 0$, so this ratio will deviate from unity.  Calculating
this involves, at one loop, the Feynman diagrams in Fig.~2, where a
dashed line stands for a light pseudoscalar propagator. Neglecting the
up and down quark masses in comparison with the strange quark mass,
this deviation has been calculated to be\cite{fiveus,goity}
\eqn{
f_{D_s} / f_D = 1 - {5\over 6}\left(1+3g^2\right) {M_K^2 \over
{16\pi^2 f^2}}\ln\left(M_K^2/ \mu^2 \right) + \lambda(\mu) M_K^2 + ...
\label{eq:ratiodecays}}
where the ellipsis denote terms with more powers of the strange quark
mass (recall $M_K^2 \sim m_s$). The dependence of $\lambda$ on the
subtraction point $\mu$ cancels that of the logarithm.  If $\mu$ is of
order the chiral symmetry breaking scale then $\lambda$ has no large
logarithms and for very small $m_s$ the explicit logarithm dominates
the deviation of $f_{D_S} / f_D$ from unity. In
Eq.~\ref{eq:ratiodecays} the contribution from $\eta$ loops has been
written in terms of $M_K$ using the Gell-Mann--Okubo formula
$M_\eta^2=4M_K^2/3$, and the contribution from pion loops,
proportional to $M_\pi^2 \ln M_\pi^2$, has been neglected.
Numerically, using $\mu=1$~GeV, the result is that
\eqn{
f_{D_s} / f_D = 1 + 0.064\ (1+3g^2),
}
or $f_{D_s} / f_D = 1.16$ for $g^2=0.5$.

\vskip1.5truecm
\epsfbox{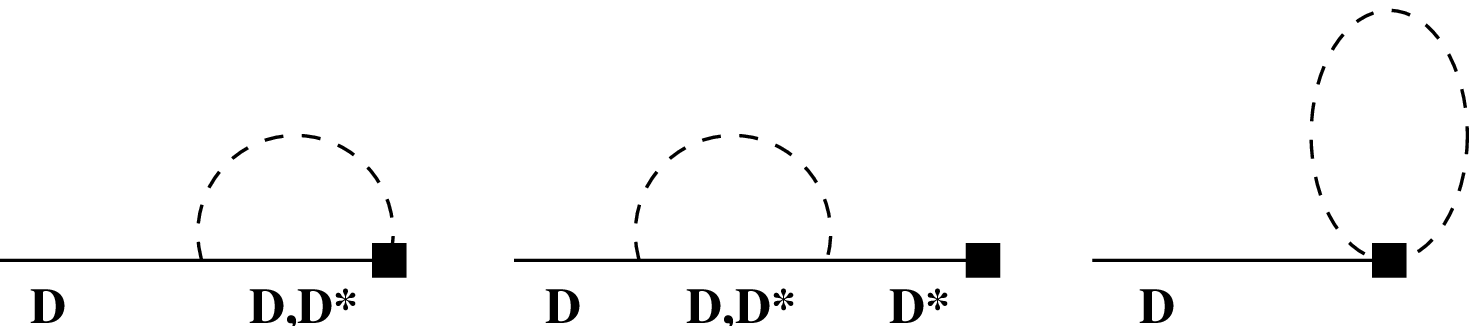}
\centerline{Figure 2. Feynman Diagrams in the calculation of $f_{D_s}
/ f_D$.}
\vskip2.0truecm

The same formula also holds for $f_{B_s}/f_B$. In fact, to leading
order in $1/M_Q$ the ratio is independent of the the flavor of the
heavy quark. Consequently,
\eqn{
{f_{B_s}/f_B \over f_{D_s} / f_D} =1  \label{eq:ratioratio}
}
to leading order in $1/M_Q$ and all orders in the light quark masses.
Now, Eq.~\ref{eq:ratioratio} also holds as a result of chiral
symmetry, for any $m_c$ and $m_b$. That is $f_{B_s}/f_B$ and
$f_{D_s}/f_D$ are separately unity in the limit in which the light
quark masses are equal. This means that  deviations from unity in
Eq.~\ref{eq:ratioratio} must be small, $O(m_s)\times
O(1/m_c-1/m_b)$.\cite{ratioofbs} This ratio of ratios is observed to
be very close to unity in a variety of calculations.\cite{soni} This
may be very useful, since it suggests obtaining the ratio
$f_{B_s}/f_B$ of interest in the analysis of $B-\bar B$ mixing (see
below) from the ratio $f_{D_s}/f_D$, measurable from leptonic $D$ and
$D_s$ decays.

The hadronic matrix elements needed for the analysis of $B-\bar B$
mixing are
\eqna{
\vev{\bar B(v)| \bar b \gamma^\mu (1-\gamma_5) d
\ \bar b \gamma^\mu (1-\gamma_5) d | B(v)} &=& {8\over 3} f_B^2 B_B ~,\\
\vev{\bar B_s(v)| \bar b \gamma^\mu (1-\gamma_5) s
\ \bar b \gamma^\mu (1-\gamma_5) s | B_s(v)}&=&
 {8\over 3} f_{B_s}^2 B_{B_s} ~,
}
where the right hand side of these equations define the parameters
$B_{B_s}$ and $B_{B}$. In the $SU(3)_V$ symmetry limit
$B_{B_s}/B_{B}=1$.  For non-zero strange quark mass, the ratio is no
longer unity.  The chiral correction is\cite{fiveus}
\eqn{
{B_{B_s} \over B_{B} } = 1 - {2\over 3}\left(1 - 3g^2\right)
{M_K^2 \over {16\pi^2 f^2}}\ln\left(M_K^2/ \mu^2 \right)~.
}
Again, $M^2_{\eta}= 4 M^2_K/3$ has been used. Using $\mu=1$~GeV,
$f=f_K$, and $g^2=0.5$, the correction is $B_{B_s} / B_{B}\approx
0.95$.

Violations to chiral symmetry in $B\to D$ semileptonic decays have
also been studied. One obtains that a different Isgur-Wise function
must be used for each flavor of light spectator quark\cite{goity}
\eqn{
\frac{{\xi_{s}(\vv)}}{{\xi_{u,d}(\vv)}}=
1+ {5\over 3}g^2\Omega(\vv) {M_K^2 \over {16\pi^2
 f^2}}\ln\left(M_K^2/ \mu^2 \right) + \lambda'(\mu,\vv) M_K^2+\cdots
}
where
\eqn{
\Omega(x) =
 -1+\frac{2+x}{2\, \sqrt{x^2-1}} \;\;
 \ln \left( \frac{x+1+\sqrt{x^2-1}}{x+1-\sqrt{x^2-1}}
\right)
+ \frac{x}{4\,\sqrt{x^{2}-1}}\;\;
\ln\left( \frac{x-\sqrt{x^2-1}}{x+\sqrt{x^2-1}} \right)
}
or, expanding about $x=1$,
\eqn{
\Omega(x)=
-\frac{1}{3}(x-1)+\frac{2}{15}(x-1)^{2}-
\frac{2}{35}(x-1)^{3}+\cdots
}
Using $g^2=0.5$ and $\mu=1$~GeV, and neglecting the counterterm one
obtains
\eqn{
\frac{{\xi_{s}(\vv)}}{{\xi_{u,d}(\vv)}}=
1- 0.21\, \Omega(\vv) +\cdots}
or a 5\% correction at $\vv=2$.

\twohead{6.4}{Violations to Heavy Quark Symmetry}

In a similar spirit one can consider the corrections in chiral
perturbation theory to predictions that follow from heavy quark spin
and flavor symmetries. These are effects that enter at order $1/M_Q$,
so the first step towards this end is to supplement the
phenomenological lagrangian with such terms. In particular, the only
$SU(3)_V$ preserving term of order $1/M_Q$ that violates spin symmetry
in the lagrangian is\cite{fiveus}
\eqn{
\Delta{\cal L}_{\rm int} =
 {\lambda_2 \over M_Q} \Tr \bar H^{(Q)a} \sigma^{\mu \nu} H_a^{(Q)}
\sigma_{\mu \nu}~. \label{eq:spinsymviol}}
In addition there are contributions to the lagrangian in order $1/M_Q$
that violate flavor but not spin symmetries. These can be
characterized as introducing $M_Q$ dependence in the couplings $g$,
$\lambda_1$ and $\lambda'_1$ of Eqs.~\ref{eq:gterm}
and~\ref{eq:lcoup}.  At the same order as these corrections, there is
a term that violates both spin and $SU(3)_V$ symmetries
\eqn{
\Delta{\cal L}_{\rm int} ={\lambda_3\over M_Q}
\Tr\left[ \bar H^{(Q)a}\sigma^{\mu \nu} H_b^{(Q)}\sigma_{\mu\nu}\right]
 m_q{}^b{}_a	\label{eq:lthree}}

Spin symmetry violation is responsible for ``hyperfine'' splittings in
spin multiplets. To leading order these mass splittings are computed
in terms of the spin symmetry violating coupling of
Eq.~\ref{eq:spinsymviol}
\eqn{
\Delta_B\equiv M_{B^*}-M_B = -{8\lambda_2\over m_b} \label{eq:hyperfsplit}}
That the mass splittings scale like $1/M_Q$ seems to be well verified
in nature:
\eqn{
{M_{D^*}-M_D \over M_{B^*}-M_B} \approx {M_B\over M_D}}

\vspace{1.5cm}

\vbox{
\centerline{Table~3. Measured Mass Splittings}
\begin{center}
\begin{tabular}{|c|c|c|}
\hline
$X-Y$ & $M_{X}-M_Y$ \\
&  (MeV)\\
\hline
$D_s - D^+$&$ 99.5 \pm 0.6$\cite{PDG}\\
 $D^+ - D^0$&$4.80 \pm 0.10 \pm 0.06  $\cite{cleomass} \\
$D^{*+} - D^{*0}$&$3.32 \pm 0.08 \pm 0.05  $\cite{cleomass}\\
$D^{*0} - D^{0}$&$142.12 \pm 0.05 \pm 0.05 $\cite{cleomass} \\
 $D^{*+} - D^{+}$ &$140.64 \pm 0.08 \pm 0.06$\cite{cleomass} \\
$D_s^{*} - D_s$&$ 141.5 \pm 1.9$\cite{PDG} \\
\hline
$ B_s - B  $&$ 82.5 \pm 2.5  $\cite{cleomass}
  or $121 \pm 9$\cite{cusb}\\
$  B^0 - B^+ $&$ 0.01 \pm 0.08  $\cite{PDG} \\
$ B^{*} - B $&$46.2 \pm 0.3 \pm 0.8$\cite{cleotwomass}
or $  45.4 \pm 1.0$\cite{cusb} \\
$  B_s^{*} - B_s $&$ 47.0 \pm 2.6 $\cite{cusb} \\
\hline
$( D^{*0} - D^0)$& \\
$-( D^{*+} - D^+ )$&$1.48 \pm 0.09 \pm 0.05 $\cite{cleomass}  \\
\hline
\end{tabular}
\end{center}
}
\vspace{1cm}

Armed with the machinery of chiral lagrangians that include both spin
and chiral symmetry violating terms, one can compare hyperfine
splitting for different flavored mesons. There is a wealth of
experimental information to draw from; see Table~3. Breaking of flavor
$SU(3)_V$ and heavy quark flavor symmetries by electromagnetic effects
is not negligible. It is readily incorporated into the lagrangian in
terms of the charge matrices $Q_Q={\rm diag}(2/3,-1/3)$ and $Q_q={\rm
diag}(2/3,-1/3,-1/3)$,\cite{rosnerwise} which must come in
bilinearly. For example, terms involving $Q_q^2$ correspond to
replacing $m_q\to Q_q$ in Eqs.~\ref{eq:lcoup} and~\ref{eq:lthree}.
The electromagnetic effects of the light quarks can be neglected if
one considers only mesons with $d$ and $s$ light quarks. The
electromagnetic shifts in the hyperfine splittings $\Delta_{X_q}$ and
$\Delta_{X_q}$ ($X=D,B$, $q=d,s$) differ on account of different $b$ and $c$
charges, but they cancel in the difference of splittings
\eqn{
 \Delta_{X_s}-\Delta_{X_d}=
(M_{X^*_s}-M_{X_s}) - (M_{X^*_d}-M_{X_d})
}
The only term in the phenomenological lagrangian that enters this
difference is Eq.~\ref{eq:lthree}. This immediately leads to
\eqn{
(M_{B^*_s}-M_{B_s}) - (M_{B^*_d}-M_{B_d}) =
(m_c / m_b)
\left({\bar\alpha_s(m_c)\over\bar\alpha_s(m_b)} \right)^{\!\!9/25}
\left[ (M_{D^*_s}-M_{D_s}) - (M_{D^*_d}-M_{D_d})\right]
 \label{eq:massrel}}
We have included here the short distance QCD effect
that is usually neglected.\cite{lukefalk}

The accuracy with which Eq.~\ref{eq:massrel} holds is to be much
better than the separate relations for each hyperfine splitting  in
Eq.~\ref{eq:hyperfsplit}. Recall that $SU(3)_V$ breaking by light
quark masses and electromagnetic interactions have been accounted for
in leading order. Moreover, the result is trivially generalized by
replacing the quark mass matrix in Eqs.~\ref{eq:lcoup}
and~\ref{eq:lthree}, by an arbitrary function of the light quark mass
matrix.  It is seen from Table~3 that this relation works well. The
left side is $1.2\pm2.7$~MeV while the right side is $3.0\pm6.3$~MeV.

Since both sides of Eq.~\ref{eq:massrel} are consistent with zero and
both are proportional to the interaction term in Eq.~\ref{eq:lthree},
it must be that the coupling $\lambda_3$ is very
small.\cite{rosnerwise} From the difference of hyperfine splittings in
the charm sector
\eqn{
-{8\lambda_3\over m_c}(m_s-m_d) = 0.9\pm1.9~{\rm MeV}}
while
\eqn{
M_{D_s}-M_{D_d}=4\lambda_1 (m_s-m_d)- {12\lambda_3\over
m_c}(m_s-m_d) = 99.5\pm0.6~{\rm MeV}}
leading to $|\lambda_3/\lambda_1|$ less than $\sim20$~MeV. This is
smaller than expected by about an order of magnitude. With such a
small coefficient it is clear that the next-to-leading terms and the
loop corrections may play an important role. In particular they may
invalidate the simple $1/M_Q$ scaling of
Eq.~\ref{eq:massrel}.\cite{ransath} There is no obvious breakdown of chiral
perturbation theory, even though the leading coupling ($\lambda_3$) is
anomalously small.\cite{jenk}

At one loop, the expressions for the mass shifts involve large
$O(m_s\ln m_s)$ and $O(m_s^{3/2})$ (non-analytic)
terms.\cite{goity,jenk} The coupling $\lambda_3$ is not anomalously
small at one loop. Instead, the smallness of the difference of
hyperfine splittings in Eq.~\ref{eq:massrel} is the result of a
precise cancellation between one loop and tree level graphs.
Explicitly,\cite{jenk}
\eqn{
\left( M_{X_s}-M_{X_s^*} \right)
-\left( M_{X_d}-M_{X_d^*} \right)  =
 \frac 5 3 g^2 \left(\frac{8\lambda_2}{M_Q}\right)
{M_K^2 \over {16 \pi^2 f^2}}
\ln \left( M_K^2/ \mu^2 \right)
-{8\lambda_3\over M_Q}m_s
}
With $g^2=0.5$ and $\mu=1$~GeV, the chiral log is 30~MeV, so the
$\lambda_3$ counterterm must cancel this to a precision of better than
10\%.

The $1/M_Q$ corrections to the masses $M_X$ and $M_{X^*}$ drop out of
the combination $M_{X}+3M_{X^*}$. The combination
$(M_{X_s}+3M_{X_s^*})-(M_{X_d}+3M_{X_d^*})$ is a measure of $SU(3)_V$
breaking by a non-vanishing $m_s$ (or $m_s-m_d$ if the $d$ quark mass
is not neglected). It can be computed in the phenomenological
lagrangian. To one loop\cite{jenk}
\eqna{
\frac 1 4 \left( M_{X_s}+3M_{X_s^*} \right)
-\frac 1 4 \left( M_{X_d}+3M_{X_d^*} \right) = 4\lambda_1 m_s
- g^2 \left( 1 + \frac 8 {3 \sqrt 3} \frac 1 2
\right) {{M_K^3} \over {16 \pi f^2}}\nonumber\\
- 4\lambda_1 m_s \left( \frac{25}{18} + \frac 9 2 g^2 \right)
{M_K^2 \over {16 \pi^2 f^2}} \ln \left( M_K^2/ \mu^2
\right)
\label{eq:spinindepmass}}
The pseudoscalar splittings $(M_{D_s}-M_{D_d})$ and
$(M_{B_s}-M_{B_d})$ have been measured; see Table~3. Also,
$\frac14(M_{X_s}+3M_{X_s^*}) -\frac14(M_{X_d}+3M_{X_d^*})=
\frac34[(M_{X^*_s}-M_{X_s})-(M_{X^*_d}-M_{X_d})] + (M_{X_s}-M_{X_d})$,
and the term in square brackets is less than a few MeV, as we saw
above. The combination $(M_{X_s}+3M_{X_s^*})-(M_{X_d}+3M_{X_d^*})$ in
Eq.~\ref{eq:spinindepmass} is first order in $m_s$ but has no
corrections at order $1/M_Q$. Thus, one expects a similar numerical
result for $B$ and $D$ systems. Experimentally,
$(M_{B_s}-M_{B_d})/(M_{D_s}-M_{D_d})$ is consistent with unity; see
Table 3. The formula in Eq.~\ref{eq:spinindepmass} has a significant
contribution from the $M_K^3$ term which is independent of the
splitting parameter $\lambda_1$.  The $M_K^3$ term gives a negative
contribution to the splitting of $\sim-250$~MeV for $g^2=0.5$.  The
chiral logarithmic correction effectively corrects the tree level
value of the parameter $\lambda_1$; for $\mu = 1$~GeV and $g^2=0.5$,
the term $4\lambda_1 m_s$ gets a correction $\approx 0.9$ times its tree
level value.  Thus, the one-loop value of $4\lambda_1m_s$ can be
significantly greater than the value determined at tree-level of
approximately $100$~MeV.

Chiral perturbation theory can be used to predict the leading
corrections to the form factors for semileptonic $B\to D$ or $D^*$
decays which are generated at low momentum, below the chiral symmetry
breaking scale. Of particular interest are corrections to the
predicted normalization of form factors at zero recoil, $\vv=1$.
According to Luke's theorem (see section~2), long distance corrections
enter first at order $1/M_Q^2$. Deviations from the predicted
normalization of form factors that arise from terms of order $1/M_Q^2$
in either the lagrangian or the current are dictated by
non-perturbative physics. But there are computable corrections that
arise from the terms of order $1/M_Q$ in the lagrangian.  These must
enter at one-loop, since Luke's theorem prevents them at tree
level, and result from the spin and flavor symmetry breaking in the
hyperfine splittings $\Delta_D$ and $\Delta_B$. Retaining only the
dependence on the larger $\Delta_D$,
the correction to the matrix elements at zero recoil are\cite{ranwise}
\eqna{
\vev{D(v) | J^{\bar c b}_\mu|B(v)} &=& 2v_\mu\left(1-
{3g^2\over2}\left({\Dc\over4\pi f}\right)^2\left[ F(\Dc/M_\pi)
+\ln(\mu^2/M_\pi^2)\right] + C(\mu)/m_c^2\right)\nonumber
\\ \label{eq:ranwisea}\\
\vev{D^*(v,\epsilon) | J^{\bar c b}_\mu|B(v)} &=&
2\epsilon^*_\mu\left(1-  {g^2\over2}\left({\Dc\over4\pi
f}\right)^2\left[ F(-\Dc/M_\pi) +\ln(\mu^2/M_\pi^2)\right] +
C'(\mu)/m_c^2\right)\nonumber
\\ \label{eq:ranwiseb} }
where $C$ and $C'$ stand for tree level counter-terms and
\eqn{
F(x)\equiv\int_0^\infty dz {z^4\over (z^2+1)^{3/2}}
\left(\frac{1}{[(z^2+1)^{1/2}+x]^2}-
\frac{1}{z^2+1}\right) \label{eq:ranwisef}}
As before, no large logarithms will appear in the functions $C$
and $C'$ if one takes $\mu\approx4\pi f\sim1$~GeV. With this
choice, formally, their contributions are dwarfed by the term that is
enhanced by a logarithm of the pion mass. Numerically, with $g^2=0.5$
the logarithmically enhanced term is $-2.1$\% and $-0.7$\% for $D$
and $D^*$, respectively.

The function $F$ accounts for effects of order $(1/m_c)^{2+n}$,
$n=1,2,\ldots$\ \ It is enhanced by powers of $1/M_\pi$ over terms that
have been neglected. Consequently it is expected to be a good
estimate of higher order $1/m_c$ corrections. With
$\Dc/M_\pi\approx1$, one needs $F(1)=14/3-2\pi$ and
$F(-1)=14/3+2\pi$ for a numerical estimate; with $\mu$ and $g^2$ as
above, this term is 0.9\% and $-2.0$\% for $D$ and $D^*$, respectively.

\twohead{6.5}{Trouble on the Horizon?}

I would like to point out a peculiar aspect of this result. The
function $F(x)$ can be expanded in $x$ starting at order $x$, as
expected.\cite{ranwise} But it can also be expanded in $1/x$, and the
leading term is a logarithmic singularity $\sim-2\ln x$. Physically
this limit corresponds to $M_\pi\to0$ (rather than the absurd
alternative $\Dc\to\infty$), and the logarithmic singularity is
canceled by the $\ln(\mu^2/M_\pi^2)$ in Eqs.~\ref{eq:ranwisea}
and~\ref{eq:ranwiseb}. Thus, the expansions in powers of $x$ and $1/x$
correspond, in terms of physical limits, to expansions in powers of
$1/m_c$ and $M_\pi$, respectively. These are alternative, but not
equivalent, expansions. This troubles me some. It seems to
indicate that the order of the limits $1/m_c\to0$ and $M_\pi\to0$
matters. But the phenomenological lagrangian for pions and heavy
mesons implicitly assumes that one can systematically expand about the
origin in $1/m_c$ -- $M_\pi$ space.

Frequently the non-analytic corrections to relations that follow from
the symmetries are uncomfortably large. A case of much interest is the
relation between the form factors $f_\pm$ and $h$ for $B\to K$
transitions, relevant to the short distance process $b\to se^+e^-$,
\eqna{
    \langle \ol K(p_K)\,|\,\ol s\gamma^\mu b\,|\,\ol B(p_B)\rangle
    &=&f_+\,(p_B+p_K)^\mu + f_-\,(p_B-p_K)^\mu\,,\\
    \langle \ol K(p_K)\,|\,\ol s\sigma^{\mu\nu} b\,|\,\ol B(p_B)\rangle
    &=& {\rm i} h\,[(p_B+p_K)^\mu (p_B-p_K)^\nu -
    (p_B+p_K)^\nu (p_B-p_K)^\mu]\,,}
and the form factors for $B\to\pi e\nu$,
\eqn{
   \langle \ol \pi(p_\pi)\,|\,\ol u\gamma^\mu b\,|\,\ol B(p_B)\rangle
    =\hat f_+\,(p_B+p_\pi)^\mu + \hat f_-\,(p_B-p_\pi)^\mu\,.}
In the combined large mass and chiral limits  only one of these form
factors is independent:
\eqn{
m_b h = f_+ = -f_- = \hat f_+ = -\hat f_- \label{eq:kpirel}}
In this limit, the ratio of rates for $B\to K e^+e^-$ and $B\to\pi e\nu$
is simply given, in the standard model of electroweak interactions, by
$|V_{ts}/V_{ub}|^2$, times a perturbatively computable function of the
top quark mass. If the relation~\ref{eq:kpirel} held to good accuracy
one could thus measure a ratio of fundamental standard model
parameters.\footnote{Another application of this
relation was discussed by I. Dunietz in this workshop.
Assuming factorization in $B\to \psi X$, ratios of CKM elements can be
extracted from these two body hadronic decays. For more details, consult
the talk by Dunietz, these proceedings.}

The non-analytic, one-loop corrections to the relations in
Eq.~\ref{eq:kpirel} have been computed.\cite{falkben} The results are
too lengthy to display here. Numerically, the violation to $SU(3)_V$
symmetry is found to be at the 40\% level.\footnote{The large
violation of $SU(3)_V$ symmetry affects as well the results of
Dunietz (see previous footnote).}

The phenomenological lagrangian that we have been considering
extensively neglects the effects of states with heavy-light quantum
numbers other than the pseudoscalar -- vector-meson multiplet. The
splitting between multiplets is of the order of 400~MeV and is hardly
negligible when one considers $SU(3)_V$ relations involving both $\pi$ and
$K$ mesons. For example, consider the effect of the scalar --
pseudovector-meson multiplet. One can incorporate its effects into the
phenomenological lagrangian. To this end, assemble its components into a
``superfield'', akin to that in Eq.~\ref{eq:superf} for the
pseudoscalar -- vector multiplet:\cite{flf}
\eqn{
S_a(v)={1+\vslash\over2}\left[
     B_{1a}^{\prime\mu}\g_\mu\g^5 - B^*_{0a}\right]\,.
                                         \label{eq:Ssuperf}}
The phenomenological lagrangian has to be supplemented with a kinetic
energy and mass for $S$,
\eqn{
  \Tr\left[\overline S_a(v)(i v\cdot D_{ba}-\Delta\delta_{ba})
			S_b(v)\right]\,,}
where $\Delta$ is the mass splitting  for the excited $S$ from the
ground state $H$, and with coupling terms
\eqn{
 g'\,\Tr\left[\ol S_a(v)S_b(v)\,\Aslash_{ba}\g^5\right] +
 (h\,\Tr\left[\ol H_a(v)S_b(v)\,\Aslash_{ba}\g^5\right] + {\rm
h.c.})\,.}
In terms of these one can now compute additional corrections to
quantities such as $f_{D_s}/f_D$ in Eq.~\ref{eq:ratiodecays}.
Numerically the corrections are not small,\cite{falk}
$ f_{D_s}/f_D = 1 + 0.13h^2 $ for $M_{D^*_0}=2300$~MeV (or
$ f_{D_s}/f_D = 1 + 0.08 h^2 $ for $M_{D^*_0}=2400$~MeV),
assuming the strange mesons to be 100~MeV heavier. Similarly,
corrections to the Isgur-Wise function can be computed, and are not
negligible.\cite{falk}

\onehead{7.}{CONCLUSIONS}
Applications of heavy quark symmetries and of heavy quark effective
theory methods abound. Many specific predictions have been made and
can be tested. If the predictions work well we may feel confident in
using these methods for a more lofty goal, that of interpreting
experiments,  be it for the measurement of fundamental parameters (as
in $|V_{cb}|$) or in probing new physics at very short distances (as
in $B\to K \ell^+\ell^-$).

Theorists are starting to understand the precision and limitations of
the method. The warning flags of the previous section are a sign of
the maturity research in this field has attained.

This is not to say the work is done. Many open questions remain. A
salient issue is that of computation of form factors for semileptonic
$b\to u$ decays. Even the inclusive rate cannot be computed at large
electron energies,\footnote{Space limitations keep me from discussing
recent work on this issue. The interested reader may consult the
literature.\cite{leptspect}} where it is measured with an aim at determining
$|V_{ub}|$. Some remaining issues require improved input from
experiment. For example, a better measurement of the entries in
Table~1 and of the lifetimes of $B^+$ and $B^0$ would settle the issue
of factorization discussed above.

Regardless of the nature of the machine that conducts the next
generation beauty and charm experiments, Heavy Quark Effective Theory
methods will play a salient role in the interpretation of the results.

\vskip1in

\onehead{8.}{ACKNOWLEDGEMENTS}

I thank Glenn Boyd for illuminating discussions. This work was supported
in part by the Department of Energy under contract number
DE-AC35-89ER40486.

\end{document}